# A SURPRISING CLARIFICATION OF THE MECHANISM OF ION-CHANNEL VOLTAGE-GATING

*AR. PL. Ashok Palaniappan*[*]

An intense controversy has surrounded the mechanism of voltage-gating in ion channels. We interpreted the two leading models of voltage-gating with respect to the thermodynamic energetics of membrane insertion of the voltage-sensing 'module' from a comprehensive set of potassium channels. KvAP is an archaeal voltage-gated potassium channel whose x-ray structure was the basis for determining the general mechanism of voltage-gating. The free energy of membrane insertion of the KvAP voltage sensor was revealed to be a single outlier. This was due to its unusual sequence that facilitated large gating movements in its native lipid membrane. This degree of free energy was the least typical of the other voltage sensors, including the Shaker potassium channel. We inferred that the two leading models of voltage-gating referred to alternative mechanisms of voltage-gating: each is applicable to an independent set of ion channels. The large motion of the voltage-sensor during gating proposed by the KvAP-paddle model of gating is unlikely to be mirrored by the majority of ion channels whose voltage sensors are not located at the membrane-cytoplasm interface in the channel closed state.

*Department of Research, Chettinad Hospital and Research Institute, Chettinad University, Kelambakkam, Tamilnadu 603013, India.
E-mail: plnppnashok@yahoo.com
Ph: +91 44 47429045

## I INTRODUCTION

With the determination of the structure of a voltage-gated K$^+$ channel[1], a controversy took shape regarding a contradiction in the principle of voltage-gating in ion channels. Two competing models of voltage-gating had been put forward, and the accumulation of evidence had done little to resolve the controversy[2]. The voltage-sensing S4 helix in all voltage-gated ion channels consists of arginine residues at intervals of third position, and the movement of these gating charges 'activates' the channel. The canonical model of voltage-gating posits that the S4, in the channel closed state, is surrounded by protein which provide an aqueous crevice to shield it from the lipid bilayer. Upon depolarisation, the S4 undergoes small vertical displacements of ~2-3Å to initiate conformational changes associated with channel opening[3,4]. The competing model of voltage-gating, called the 'paddle model,' stipulates that the S4, in the channel closed state, is located at the membrane-cytoplasm interface, and upon depolarization, undergoes a large transverse movement of ~20Å to translocate to the extracellular side and gate the channel[5,6,7].

## 2 METHODS

The paddle model is supported primarily by studies with the KvAP channel in its native lipid membrane. Experimental studies with other voltage-gated K$^+$ channels, especially the eukaryotic Shaker channel, have unambiguously supported the alternative 'canonical' model. Shrivastava et al considered whether the KvAP channel is 'different' from other voltage-gated K$^+$ channels[8] and concluded otherwise based on the conservation of the essential S4 motif of basic residues.

Hessa et al developed an amino acid hydrophobicity scale[9]. Using this hydrophobicity scale, the free energy (ΔG) of membrane insertion of the KvAP voltage sensor was determined to be ~0 kcal/mol[10]. This provided the necessary thermodynamic basis for the location of the KvAP voltage sensor at the membrane-cytoplasm interface in the channel closed state. Here, we extend the investigation of thermodynamic stability of the voltage sensor to more contentious sequences of voltage sensors. We constructed a comprehensive dataset of 147 90% non-redundant voltage-gated K$^+$ channels representing various subfamilies including KCNA (Shaker), KCNB (Shab), KCNC (Shaw), KCND (Shal), KCNF, KCNG, and KCNS. The ΔG of membrane insertion of the voltage sensor of each channel was scored using the amino acid hydrophobicity scale.

## 3 RESULTS AND DISCUSSION

Fig. 1 shows the ΔG of membrane insertion for the voltage sensors in the comprehensive dataset. It was seen that all the S4's possessed a significant positive free energy, which would in turn function to impede all large motions in the lipid environment of the membrane. Therefore it is unlikely that the voltage sensors were located at the membrane-cytoplasm interface in the channel closed state, since gating would entail large movements of the voltage sensor in the membrane for such a location. This result provided support for the 'canonical' model of gating for the majority of voltage sensors.

On the other hand, the KvAP S4 voltage sensor was the clear outlier in the plot, with its ΔG of membrane insertion more than 3σ from the mean ΔG. This finding complicated the extension of conclusions about voltage-gating drawn from studies of the KvAP channel. The paddle model of gating applied to only those channels with an 'interfacial' S4 in the closed state, which is implausible for the majority of channels. It is worth noting that the ΔG of membrane insertion of the voltage sensors of channels homologous to the KCND subfamily was as high as 16 kcal/mol.

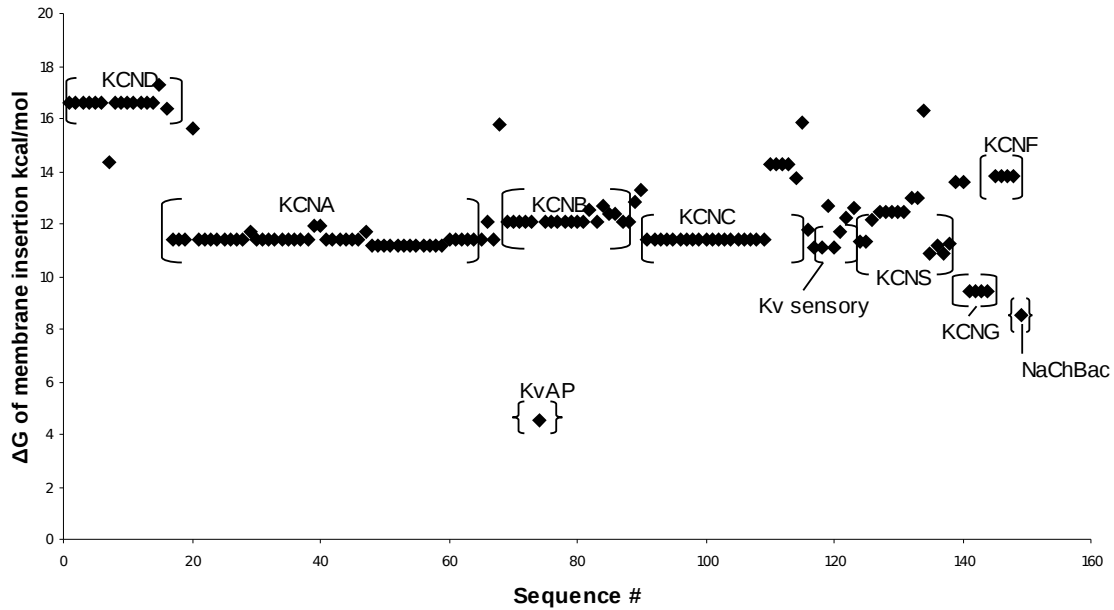

**Figure 1** ΔG of membrane insertion of voltage sensors of a comprehensive (prokaryotic and eukaryotic) set of 147 90% non-redundant K$^+$-channels, grouped by homology to the closest eukaryotic subfamily. KvAP' has the lowest insertional ΔG. After correction for position-dependence, the KvAP value is close to zero[10].

The voltage-gating properties of the prokaryotic one-domain sodium channel NaChBac might be similar to the KvAP channel[11]. Figure 1 indicates the ΔG of membrane insertion of the NaChBac voltage sensor, which is midway between the KvAP voltage sensor and the rest of the voltage sensors.

Fig. 2 shows the residue frequencies in the same comprehensive set of channels compared with the KvAP S4. It was seen that seven of the 18 residues in the KvAP S4 were leucines, whereas the comprehensive set had about three leucines on average. A variety of less hydrophobic residues such as M, A, T, H, and K were seen in the universal group but absent in the KvAP sequence. Also shown in fig. 2 is the alignment of the consensus sequence of the comprehensive set with the KvAP and

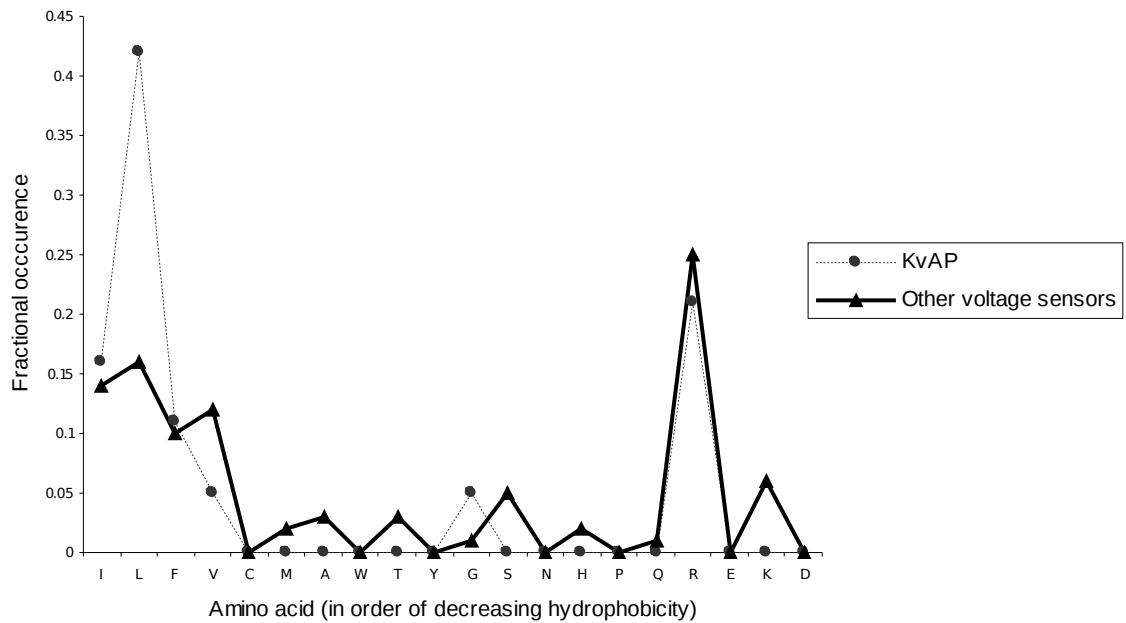

```
S4Consensus    ILRVIRLVRIFRIFRLSR
KvAP           LFRLVRLLRFLRILLIIS
NaChBac        VIRILRVLRVLRAISVVP
```

**Figure 2** Residue frequencies of the comprehensive set vs KvAP. There is a mismatch in the propensity for hydrophobicity between the comprehensive set and KvAP. Below, alignment of the consensus S4 sequence of the comprehensive set with the KvAP and NaChBac S4 sequences. Arginines are highlighted.

NaChBac S4 sequences. The consensus sequence differed from the KvAP sequence in a total of 12 out of the 18 positions. On the contrary, the consensus sequence differed from the Shaker S4 sequence in only one position (position #10: I↔V). (See fig. 3 for comprehensive alignment.)

## 4   CONCLUSION

The ΔG of membrane insertion of the voltage sensor constrains the magnitude of its translocation in the phospholipid membrane during gating. The higher the ΔG, the greater is the constraint for movement. The KvAP voltage sensor has the lowest ΔG, and is therefore capable of maximum translocation. As ΔG increases, there is a continuum of possible translocations of decreasing magnitude. For the majority of voltage sensors, free interaction with lipid is energetically unfavourable. Hence we would expect that they undergo minimal translocation during gating, whereas the paddle model of gating required the umimpeded interaction between the phospholipid and ion channel in both the channel open and closed states. Later results from Dr. R. MacKinnon's laboratory showed that the phospholipid membrane played a role in KvAP gating[12]. This implied that there is co-evolution between the voltage sensor of the ion channel and the phospholipid membrane. Co-evolution is emerging as an important player in biology; there is substantial evidence for co-evolution between catalysis and regulation in potassium channels in particular[13]. More work is required to ascertain the chemical interactions between ion-channel and membrane, and their correspondence with the exhibited mechanism of gating. It would be interesting to explore if the existence of multiple mechanisms for gating is an exceptional observation in the study of protein function.

```
 1 1ORQ_C       100.0%   LFRLVRLLRFLRILLIIS
 2 41351867      16.7%   TLRVFRVFRIFKFSRHSQ
 3 1763619       16.7%   SLRVFRVFRIFKFSRHSK
 4 1763617       27.8%   AFRSVRIIRVFKLARHSQ
 5 50731321      33.3%   ILRVIRLVRVFRIFKLSR
 6 47212272      27.8%   VIRLVRVFRIFKLSRHSK
 7 25815097      27.8%   VLRVLRLVRVFRVFKLSR
 8 987511        33.3%   VLRVVRVIRVIRIFKLTR
 9 987509        22.2%   MLRVIRVLRVFKLSRHSR
10 55653032      33.3%   IFRIMRILRILKLARHST
11 2315214       33.3%   VFRIMRVLRILKLARHST
12 29373793      33.3%   AIRTLRGLRIIRVLRMFK
13 3387822       22.2%   FLRVIRLVRVFKLTKHST
14 3023480       38.9%   FLRVVRFVRILRIFKLTR
15 28574020      22.2%   IIRIMRLFKLTRHSSGLK
16 3219511       22.2%   IIRIMRLCKLTRHSAGLK
17 45550160      22.2%   IVRIMRLFKLTRHSPGLR
18 50057616      27.8%   GLRVIRIIRFMRVFRLFK
19 32564679      33.3%   VIRILRVLRVIRVLKLGR
20 25148255      38.9%   TVRLLRVLRVIRIAKLGR
21 39593041      33.3%   VVRILRVLRVIRIIKLGR
22 38638831      38.9%   VVRVLRVLRVVRILKLGR
23 39593393      27.8%   VVRVLRVLRMARVFKLAR
24 34146982      33.3%   VVRVMRLARVARIFKLAR
25 54697186      27.8%   ILRLMRIFRILKLARHSV
26 47210943      22.2%   VLRLMRVFRIFKLARHSV
27 50510805      27.8%   VLRLMRIFRILKLARHST
28 55652626      33.3%   VFRLMRIFRVLKLARHST
29 55962430      27.8%   VLRLMRSFRVLKLARHSE
30 34147232      38.9%   VLRIMRLMRIFRILKLAR
31 47216576      33.3%   VLKVVKLMRIFRILKLAR
32 57095416      33.3%   VLRLLRALRMLKLGRHST
33 6006605       33.3%   VLRLLRALRVLYVMRLAR
34 51712737      27.8%   VLRVLRALRILYVMRLAR
35 34863358      22.2%   ALRIMRIARIFKLARHSS
```

**Figure 3** Alignment of 90% non-redundant S4 sequences from the comprehensive dataset. Accession codes refer to the Genbank Gene Products database. At the top is KvAP, and the percent values denote homology to the KvAP sequence in terms of sequence identity. The degree of homology between the KvAP S4 and all the other S4 sequences is <40% in every case, with only 17% homologies also observed.